# Triplet excitation and electroluminescence from a supramolecular monolayer embedded in a boron nitride tunnel barrier

Simon A. Svatek[1*†], James Kerfoot[1†], Alex Summerfield[1], Anton S. Nizovtsev[2,3], Vladimir V. Korolkov[1], Takashi Taniguchi[4], Kenji Watanabe[4], Elisa Antolín[5], Elena Besley[2] and Peter H. Beton[1*]

1. School of Physics and Astronomy, University of Nottingham, Nottingham, NG7 2RD, UK
2. School of Chemistry, University of Nottingham, Nottingham, NG7 2RD, UK
3. Nikolaev Institute of Inorganic Chemistry, Siberian Branch of the Russian Academy of Sciences, Academician Lavrentiev Avenue 3, 630090, Novosibirsk, Russian Federation
4. National Institute for Materials Science, 1-1 Namiki, Tsukuba, Ibaraki 305-0044, Japan.
5. Universidad Politécnica de Madrid - Instituto de Energía Solar, Avenida Complutense 30, 28040 Madrid, Spain

Abstract

We show that ordered monolayers of organic molecules stabilized by hydrogen bonding on the surface of exfoliated few-layer hexagonal boron nitride (hBN) flakes may be incorporated into van der Waals heterostructures with integral few-layer graphene contacts forming a molecular/2D hybrid tunneling diode. Electrons can tunnel from through the hBN/molecular barrier under an applied voltage $V_{SD}$ and we observe molecular electroluminescence from an excited singlet state with an emitted photon energy $h\nu > eV_{SD}$, indicating up-conversion by energies up to ~ 1 eV. We show that tunnelling electrons excite embedded molecules into singlet states in a two-step process via an intermediate triplet state through inelastic scattering and also observe direct emission from the triplet state. These heterostructures provide a solid-state device in which spin-triplet states, which cannot be generated by optical transitions, can be controllably excited and provide a new route to investigate the physics, chemistry and quantum spin-based applications of triplet generation, emission and molecular photon up-conversion.

Two-dimensional supramolecular arrays stabilised by non-covalent interactions provide a highly flexible route to the spatial organization, down to the molecular scale, of functional molecules on a surface[1–4]. While this route to surface patterning has been successful in positioning chemical groups within adsorbed monolayers, the non-covalent nature of the stabilising interactions has limited the possibilities to explore the transport of charge through the component molecules. One possible route to explore the electrical properties of supramolecular monolayers is through the use of adjacent charge injection layers placed above and/or below the molecules. This architecture, in which current flows perpendicular to the plane of the adsorbed molecules, would require that the supramolecular layer is embedded in a more complex heterostructure with integral contact and spacer layers. Although there has been recent progress in the growth of all-organic epitaxial supramolecular heterojunctions[5–7], these structures cannot currently be prepared with the required complexity and control. The techniques used to fabricate van der Waals heterostructures[8], such as a tunnel diode formed by placing few-layer hexagonal boron nitride (hBN) between two graphene layers[9], offer an alternative approach. In this Letter we show that a similar device architecture may be employed to embed a supramolecular monolayer between two hBN tunnel barriers, thus forming a hybrid molecular/2D device. The encapsulated organic molecules can be excited electrically and subsequently relax through the emission of photons resulting in electroluminescence from both singlet and spin-triplet states. Photons are up-converted by energies up to 1 eV, and we show that singlets are excited through a multi-electron inelastic process via a triplet intermediate state. This hybrid structure provides a solid-state device in which triplets can be controllably excited, and offers a route to fundamental studies of long-lived excitations with non-zero spin and their relevance to low voltage light emitting devices, and quantum spin-based excitonic and electronic devices.

To fabricate our devices, we use polymer stamp-assisted van der Waals assembly[10–12] to sequentially pick up flakes of few-layer graphene (FLG) and hBN. hBN flakes with adsorbed monolayers of organic molecules can be picked up and deposited as part of the assembly process in the same way as pristine flakes, thus allowing the integration of molecular layers within van der Waals heterostructures. We first use a polymer 'stamp' to pick up a large hBN flake (lateral dimensions 10s of μm, thickness 10s of nm), which is ultimately used to cap the device. This hBN flake is then used to pick up a FLG flake which serves as the top contact, followed by a thin (≤ 1 nm) hBN flake which forms the upper tunnel barrier. The van der Waals stack is then used to pick up a second hBN tunnel barrier, also with a thickness of 1-3 monolayers, on which a molecular monolayer has been deposited by sublimation. This part-formed tunnelling device is then released from the stamp onto a second FLG flake which forms the lower contact; the release site is chosen so that the upper and lower FLG layers make independent contact with two pre-formed contacts (10 nm Cr/30 nm Au). Further details are provided in Supporting Information (SI).

A schematic of a completed device with an embedded monolayer of sublimed perylene tetracarboxylic di-imide (PTCDI) is shown in Figure 1a. The FLG provides semi-transparent top- and bottom-electrodes and the hBN layers allow carrier injection via tunnelling under an applied bias, while suppressing quenching from the FLG contacts[13]. PTCDI is a planar molecule and is adsorbed parallel to the hBN substrate in two-dimensional islands stabilised by hydrogen bonding[14]. The islands have monolayer height, typical lateral dimensions of 5 – 10 μm and, for the deposition parameters we use (see SI), a surface coverage of 50%. The molecular ordering within the islands is resolved using atomic force microscopy (AFM) which reveal lattice vectors close to the expected value[14] (see Fig. 1b). At a large scale, individual monolayer PTCDI islands may be identified using optical microscopy during the transfer process (Fig. 1c). Images of a completed device are shown in Fig. 1d; the active area of the device (where the upper and lower FLG layers overlap) is marked. It is possible to selectively encapsulate a selected monolayer-height PTCDI island in the active region; for this device we used the island highlighted by an arrow in Fig. 1c which had been adsorbed on a bilayer hBN flake which forms the lower tunnel barrier. The effective area of the device is estimated to be 4 μm$^2$. The deposition of PTCDI, and the preparation of the hBN surface is discussed in SI.

The current-voltage characteristics of the device were measured in an optical cryostat at a temperature T = 6 ± 1 K, and are highly non-linear as expected for a tunnelling device (see Fig. 1f). The devices emit light when a current flows. The electroluminescence (EL) spectrum acquired for an applied bias $V_{SD}$ = -3.3 V (Fig. 1e) shows an intense peak at a wavelength 586.1 ± 0.5 nm (2.115 ± 0.002 eV) corresponding to the zero phonon transition (0-0) from the lowest excited singlet, $S_1$, to the ground state, $S_0$, accompanied by a satellite vibronic 0-1 peak (633.4 ± 0.5 nm/1.958 ± 0.002 eV). An optical image of the device under bias confirms that the active region of the device is the source of the photon emission (Fig. 1d). EL emission is observed in both polarities and the peak position varies by less than 0.5 nm over the measured voltage range (see Figure 2).

The EL peaks are close to the 0-0 transition (588.1 ± 0.5 nm/2.108 ± 0.002 eV) in the photoluminescence (PL) spectrum of the device (Fig. 1e). However both the PL and EL peaks are shifted from the 0-0 transition (565.0 ± 0.5 nm/2.194 ± 0.002 eV) measured for uncapped PTCDI (also shown in Fig. 1e) on a single hBN layer by 0.086 ± 0.003 eV (EL and PL spectra are also readily acquired at room temperature; the PL peak positions are independent of voltage – see SI). We have previously shown that, for uncapped PTCDI on hBN, a combination of resonant and non-resonant interactions between adsorbed PTCDI and hBN leads to a red-shift of 0.31 eV compared with gas phase spectra[15,16]. A further red-shift is expected when a second hBN layer is added in the encapsulation process due to the additional changes in dielectric environment. This is discussed in more detail in SI where we calculate the transition energies of uncapped (2.26 eV) and capped (2.20 eV) PTCDI, and, in particular, a predicted red shift, 0.06 eV, due to the addition of a second hBN layer; these values are in good agreement with our experimental values.

The EL spectrum of PTCDI exhibits significant up-conversion, that is emission of photons with energies $h\nu > eV_{SD}$, the energy gained by a tunnelling electron when passing ballistically between the two contacts[17]. This implies that the emission must occur through a multi-electron process, rather than through simple charge injection into the HOMO and LUMO levels of the molecule (as might be expected for a molecular analogue of a van der Waals heterostructure with a transition metal dichalcogenide emissive layer[13]), and is suggestive of an inelastic scattering mechanism. Figure 3a shows spectra acquired in this voltage range and we see a significant intensity in the 0-0 peak down to $|V_{SD}|$ = 1.6 V, and at room temperature (see SI) down to 1.3 V implying an up-conversion energy much greater than $k_BT$ (ruling out thermal effects) and close to 1 eV. In this device we also resolve an additional weak peak at $543.8 \pm 0.5$ nm ($2.280 \pm 0.002$ eV). We tentatively assign this feature to a 1-0 vibronic transition, although the possible origin of a PTCDI in a different trapped conformation cannot be ruled out.

The observation of photon up-conversion indicates that molecules are excited into an intermediate state and then, through a further excitation, are excited into the singlet excited state $S_1$. Electroluminescence is observed in this device for a current density down to ~ 1 pA nm$^{-2}$, which corresponds, approximately, to an average time between the traversal of electrons through each molecule of order 0.1 μs (assuming ~ 1 nm$^2$/molecule). The relevant excitation for the above-threshold emission must have a lifetime of this order of magnitude (or longer) thus ruling out mechanisms such as the sequential excitation of vibronic modes of the $S_0$ electronic ground state which relax on a much more rapid timescale. In common with Chen et al.[18] we suggest that the long lifetime strongly suggests that the intermediate state involved in photon up-conversion is a $T_1$ spin-triplet.

The role of triplets is confirmed through the identification of an additional peak in the EL spectrum at ~$1004 \pm 1$ nm, photon energy $1.235 \pm 0.001$ eV. This peak is not present in the PL spectrum (see Fig. 3a inset), but is close to the value, 1.18 eV, of the triplet state of a related PTCDI derivative determined[19] by triplet pair absorption. Accordingly we attribute this peak to emission from the $T_1$ state. The observed energy is also close to the energy, 1.29 eV, calculated for this transition (see SI).

Figure 3b shows that the emission intensities $I(T_1)$ and $I(S_1)$ from, respectively, the $T_1$ and $S_1$ states have a highly non-linear dependence on current. A logarithmic plot (Fig. 3c) shows a power law dependence of $I(S_1)$ on $I(T_1)$ over a large voltage range in both forward and reverse bias, $I(S_1) \propto I(T_1)^k$ where $k \approx 1.2$ indicating a near linear relationship. Assuming that, for a given applied voltage, the number of excited triplets is proportional to the $T_1$ intensity, triplet-triplet annihilation (TTA)[20] may be ruled out as a route to the secondary excitation of molecules from $T_1$ to $S_1$, since for this mechanism a quadratic dependence[21] ($k \geq 2$) on the number of triplets would be expected (assuming a low density of triplets[22] which is likely for the low currents we observe close to the threshold for emission). We therefore suggest that the $T_1$ to $S_1$ transition is promoted by a second inelastic electron scattering event.

The proposed mechanism is summarised in the band diagrams shown in Figures 4a and b. No emission is expected until the energy gained by tunnelling electrons, $eV_{SD}$ exceeds the energy difference between the $S_0$ and $T_1$ states thus permitting excitation between these states through an inelastic process. We also see evidence for inelastic scattering in the electrical characteristics of the device. Figure 4c shows that a broad peak is observed in the second derivative of $d_2I/dV_2$ at ~ 1.1 V, very close to the triplet energy. Peaks in $d_2I/dV_2$ due to inelastic electron scattering are expected when the voltage drop matches the energy of an excitation to which electrons are coupled[23,24] (similar features in hBN/graphene tunnel devices are observed to inelastic scattering of phonons[25]). Molecules in the $T_1$ state may undergo a further inelastic excitation to the $S_1$ state or relax via the emission of a photon. This simple model is consistent with the voltage thresholds, peak energies and variation of intensity ratio which we observe and implies that the inelastic scattering process induces a change in the spin state of the molecule.

The generation of excitons in our devices shows fundamental differences to the mechanism in conventional organic light emitting diodes (OLEDs) where electrons and holes are injected from remote contacts and combine to form both singlet and triplet excitons. The spin degeneracy of these states typically leads[26] to a relative population of singlets and triplets in a ratio 1:3. The close proximity of the charge injection layers (~ 1 nm) to the emissive monolayer in our devices provides a different route which leads to the selective excitation of spin triplets.

While the architecture of our devices is significantly different to conventional OLEDs, we draw analogies with luminescence generated by the tip of a scanning tunnelling microscope[27,28] (STML) for which the electrodes and photon source are also in close proximity and, moreover, the widely-accepted mechanism for molecular excitation in STML is via inelastic scattering and there are reports of photon up-conversion[29–31]. Furthermore, the efficiency of our devices, typically $10^{-6}$- $10^{-8}$ photons per electron is similar to that observed in STML experiments[27]. The coupling of tip-induced localised plasmons to an adsorbed molecule is considered to be the most significant STML emission process[27]; this is unlikely to be significant in our devices since the plasmon energy is much smaller[32] (in the range of 100s of meV) for our FLG contacts. However, recent papers have highlighted the role of triplets in STML. Specifically $T_1$ emission[33] has been observed in the STML from perylene tetracarboxylic dianhydride (PTCDA), although the proposed mechanism involves a charged molecule and is not accompanied by up-conversion. A triplet-mediated process has been proposed for the up-conversion in the STML of metal-free phthalocyanine[18]; this mechanism is similar to that discussed above but it is not accompanied by triplet emission.

These molecular/2D hybrid heterostructures provide an alternative electronic method to control the excitation of triplets, a transition which is optically forbidden, and offer a route to fundamental studies of long-lived optical excitations and their applications in quantum and spin based optoelectronics, and are also relevant to low voltage light emitting devices. The device architecture has many possible variations in the choice of contact materials, molecules with higher/lower energy levels, tunnel barrier width to control the current density, as well as the scaling of the active region – either down to a single or small ensemble of molecules, or up to large area device incorporating a complete monolayer. We also envisage the integration of molecular layers with more complex multilayer and/or in-plane supramolecular order to control the coupling of neighbouring molecular emitters. These solid state devices thus provide a new route to the investigation of the physics, chemistry and optoelectronic applications of triplet generation, emission and up-conversion.

ASSOCIATED CONTENT

**Supporting Information**

The Supporting Information is available free of charge on the ACS Publications website at DOI:

*to be confirmed*.

AUTHOR INFORMATION

**Corresponding Authors**

*e-mail: simon.svatek@upm.es and peter.beton@nottingham.ac.uk

**Author Contributions**

†These authors made equal contributions to this work.

ACKNOWLEDGMENTS

This work was supported by the Engineering and Physical Sciences Research Council [grant number EP/N033906/1]; and the Leverhulme Trust [grant number RPG-2016-104]. K.W. and T.T. acknowledge support from the Elemental Strategy Initiative conducted by the MEXT, Japan and the CREST (JPMJCR15F3), JST. E.A. is grateful for the Ramón y Cajal Fellowship funded by the Spanish MINECO (RYC-2015-18539). Computations were performed at the High Performance Computing facility at the University of Nottingham. We acknowledge the use of Athena at HPC Midlands+, which was funded by the EPSRC on grant EP/P020232/1 as part of the HPC Midlands+ consortium. We thank Bjarke Sørensen Jessen for helpful discussions.

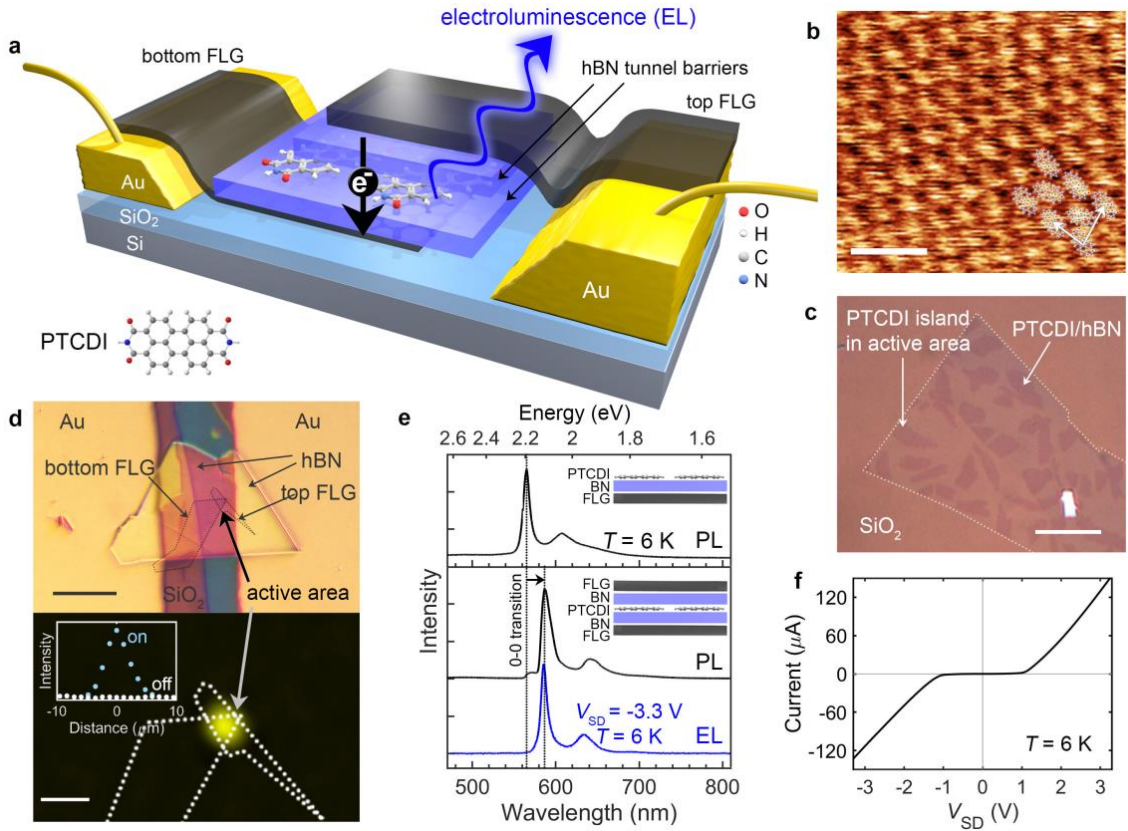

Figure 1. FLG/BN/PTCDI/hBN/FLG heterostructures. **a,** Schematic of a device in which a monolayer of PTCDI is encapsulated between two hBN tunnel barriers and charge is injected from upper and lower FLG contacts (the upper thick hBN layer, and a supporting thick hBN flake which provides a supporting substrate for the lower graphene are omitted for clarity; neither plays an active role in the device operation); lower left - schematic of the molecular structure of PTCDI. **b,** AFM image of a monolayer-thick island of PTCDI on hBN; the lattice vectors of the molecular array are marked and have the following values: 1.48 ± 0.1 nm and 1.78 ± 0.1 nm, subtended by an angle of 89°. **c,** Optical micrograph showing monolayer islands following sublimation of 0.5 monolayers of PTCDI on a bilayer hBN flake (highlighted by dotted outline); this flake forms the lower tunnel barrier for this device and the island selected for the active region is marked by an arrow – note that the PTCDI grows in a different morphology on the surrounding $SiO_2$ surface and cannot be resolved in these regions. **d,** upper - Optical image of a device showing gold contacts and van der Waals heterostructure. The active area of the device where the upper and lower FLG layers overlap is highlighted; lower - optical image of the device taken under bias ($V_{SD}$ = -3.2 V) acquired with an exposure time of 8 s with an overlay also showing the position of the graphene contacts and confirming that light is emitted from the active area of the device; inset the variation of intensity along the horizontal axis through (on) and away (off) the active region. **e,** Electroluminescence (acquisition time 100 s) and photoluminescence spectra of device acquired at liquid helium temperatures; the PL for an uncapped device is also included highlighting the peak shift due to encapsulation. **f,** Current–voltage characteristics. Scale bars (b) 3nm, (c) 20 μm, (d) 20 μm (upper) and 6 μm (lower).

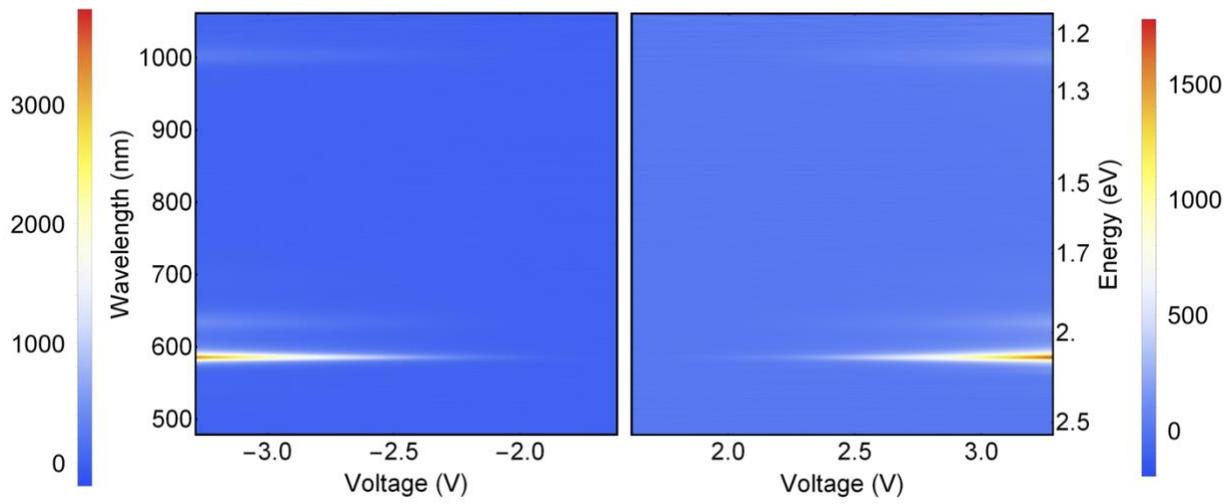

Figure 2. Voltage map of electroluminescence. The EL intensity is displayed as a colour map for different voltages (horizontal axis) and wavelengths (vertical axis). This data shows that the EL peak position is constant over the measured voltage range.

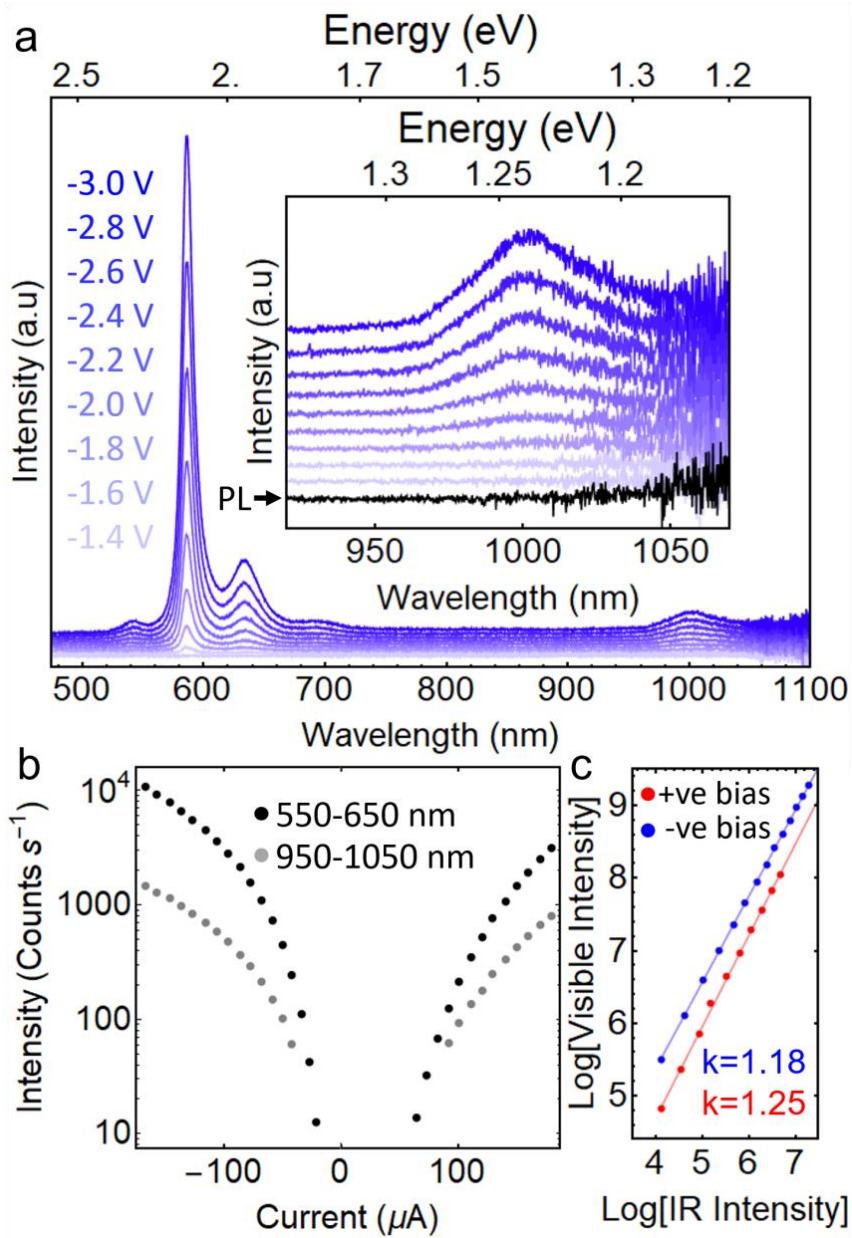

Figure 3. Photon up-conversion and triplet electroluminescence from a FLG/BN/PTCDI/hBN/FLG heterostructure. **a,** Spectra acquired at 6 K with a 100 second integration time for a series of applied voltages ranging from -1.4V to -3.0 V; inset - electroluminescence in the infra-red plotted together with the photoluminescence spectrum in the same spectral region (black). **b,** The electroluminescence signal (with background subtracted) integrated between 550 and 650 nm and 950 and 1050 nm, versus current. **c,** Logarithmic plot of the integrated intensity between 550 and 650 nm versus the intensity between 950 and 1050 nm showing a near-linear dependence between the singlet and triplet emission intensities.

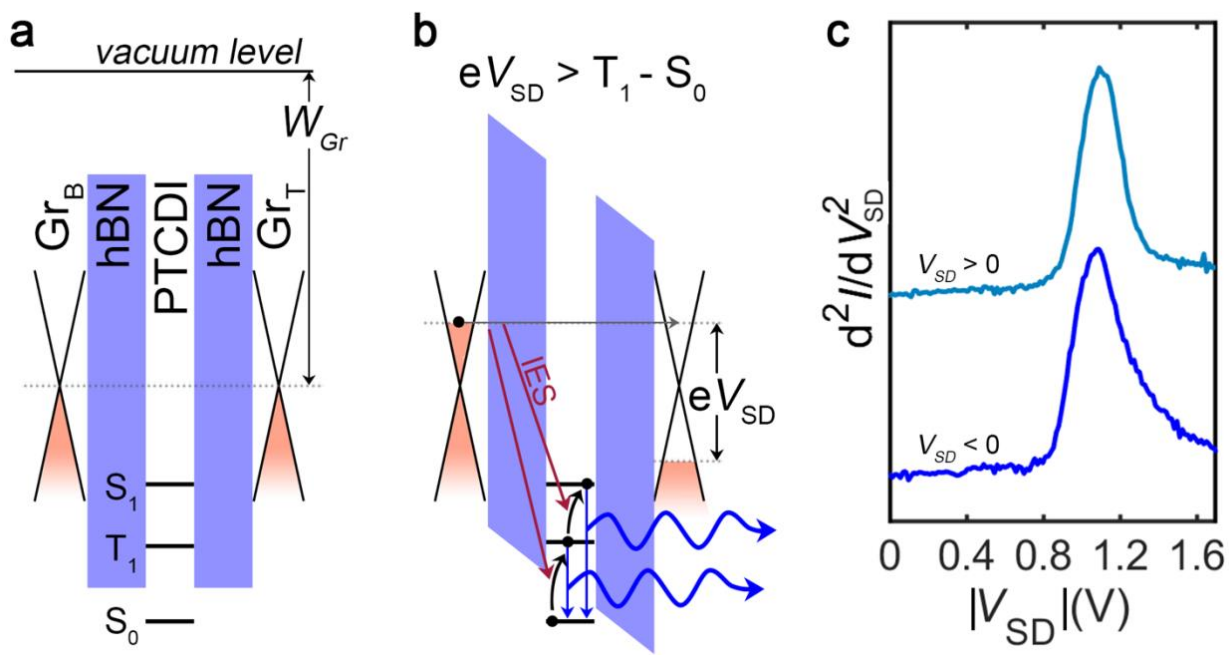

Figure 4. Mechanism and inelastic tunnelling spectroscopy. **a,** Alignment of work function of graphene ($W_{Gr}$), hBN bands and molecular energy levels under zero bias. **b,** Band alignment under bias $V_{SD}$; molecule can be inelastically excited when $eV_{SD}$ exceeds the energy difference between the $S_0$ and $T_1$ states. Molecule undergoes a further inelastic excitation to the $S_1$ state; photons can be emitted from a transition to $S_0$ from either the $S_1$ or $T_1$ states. **c,** Peaks in $d_2I/dV_2$ occur where $eV_{SD}$ matches the energy of the excitation. Clear peaks are observed for $V_{SD} \approx 1.1$ V close to the value expected for triplet excitation via inelastic scattering.